\documentclass[usegraphicx,usenatbib]{mn2e}

\title{Self-Gravity Driven Instabilities of  Interfaces in
the ISM}
\author[R.M. Hueckstaedt el al.]
    {R.M. Hueckstaedt,$^1$\thanks{E-mail:rmhx@lanl.gov}
     J.H. Hunter, Jr.,$^2$ and R.V.E. Lovelace$^3$ \\
    $^1$ Applied Physics Division, Los Alamos National Laboratory,
              Los Alamos, NM 87545, USA \\
     $^2$ Department of Astronomy, University of Florida,
             Gainesville, FL 32611, USA \\
     $^3$ Departments of Applied Physics and Astronomy, Cornell University,
                Ithaca, NY 14853, USA}

\date{\today}
\pagerange{\pageref{firstpage}--\pageref{lastpage}}
\pubyear{2006}

\begin{document}

\label{firstpage}

\maketitle

\begin{abstract}
In order to understand star formation it is important to understand the
dynamics of atomic and molecular clouds in the interstellar medium (ISM).
Nonlinear hydrodynamic flows are a key component to the ISM.
One route by which nonlinear flows arise is the onset and evolution of interfacial
 instabilities. Interfacial instabilities act to modify the interface between gas
components at different densities and temperatures.
Such an interface may be subject to a host of instabilities,
including the Rayleigh-Taylor, Kelvin-Helmholtz, and Richtmyer-Meshkov
instabilities. Recently, a new density interface instability was identified.
This self-gravity interfacial instability (SGI) causes any displacement of the interface to grow on
roughly a free-fall time scale, even when the perturbation wavelength
is much less than the Jeans length. In previous work, we used numerical
simulations to confirm the expectations of linear theory and examine the
nonlinear evolution of the SGI. We now continue our study by
generalizing our initial conditions to allow the acceleration due to self-gravity
to be non-zero across the interface. We also consider the behaviour of the SGI for
perturbation wavelengths near the Jeans wavelength.
We conclude that the action of self-gravity across a density interface may play a significant
role in the ISM either by fueling the growth of new instabilities or modifying
the evolution of existing instabilities.
\end{abstract}


\section{Introduction}
Hydrodynamic instabilities of interfaces in the interstellar medium (ISM)
have received renewed attention in part due the remarkable Hubble  images of
``elephant trunks'' and ``pillars'' in the Eagle Nebula (Hester et  al. 1996; Pound
1998). Most of the theoretical and simulation work has focused on the
stability/instability of ionization fronts driven by
UV radiation of nearby OB stars (e.g., Mizuta et al. 2005; Williams 2002).
 The fronts  may undergo acceleration so as to give the
classical Rayleigh-Taylor instability but the influence of
recombination may suppress the instability (Mizuta et al. 2005).

Hunter, Whitaker, and Lovelace (1997, 1998; hereafter Papers 1 and 2)
studied the stability of interfaces in more quiescent regions
of the ISM where the ionizing radiation is not important.
They identified a new interfacial instability which is driven
by self-gravity and acts at a density discontinuity.
This self-gravity interfacial instability (SGI) persists in the
static limit for all wavelengths and occurs in addition to
the classical Rayleigh-Taylor instability. Using a normal mode analysis, they 
derived the linear growth rate of the SGI in compressible media in relative motion
(allowing for the influence of Kelvin-Helmholtz instability).
In the incompressible limit, the growth rate for a perturbation
$\sim \exp(-i\omega t)$ of a planar interface is
\begin{equation}\label{rates}
\omega^2=\frac{-2\pi{G}(\rho_2-\rho_1)^2}{\rho_2+\rho_1}+
          \frac{gk(\rho_2-\rho_1)}{\rho_2+\rho_1}~.
\end{equation}
In equation (\ref{rates}), $k$ is the horizontal perturbation wave number ($k>0$),
$g$ a constant background acceleration, and $G$ the gravitational constant.
The mass densities of the lower and upper fluids are specified as
$\rho_{1}$ and $\rho_{2}$, respectively. The second term gives the growth rate
for the incompressible Rayleigh-Taylor (RT) instability,
and the first term is the incompressible growth rate for the SGI.  
Both instabilities persist in the static limit, but
several important differences are evident. Self-gravity knows no preferred
direction, so the SGI is destabilizing across any density interface. An
interface is RT unstable only if $g(\rho_{2}-\rho_{1})<0$, such that the
heavy fluid sits ``on top'' of the light fluid. The SGI growth rate depends upon
the absolute densities of the fluids and their ratio but not the perturbation
wavelength. The RT growth rate changes with perturbation wavelength and density
ratio, but it does not depend upon the absolute densities in the fluids.
Given these dependencies, the SGI is expected to grow faster than the RT
instability for a fixed value of $g$ when the perturbation wavelength is
long enough such that $\lambda=2\pi/k>g/G\vert\rho_{2}-\rho_{1}\vert.$
In planar geometry, the growth rate for the SGI depends only weakly upon the Jeans 
criterion for the fully compressible case and not at all upon the perturbation 
wavelength in the incompressible limit. The underlying reason that self-gravity is able 
to drive an instability for any wave number is that the  configuration is not one 
of minimum energy.

Numerical studies contrasting the behaviours of the SGI and RT instabilities
have been performed by Hueckstaedt and Hunter (2001; Paper 3) and by
Hueckstaedt, Peterson, and Hunter (2005; Paper 4). In the nonlinear regime, the
SGI evolves such that the growth of tenuous bubbles outpaces that of dense
spikes; whereas, the RT instability is characterized by dense spikes streaming
into the tenuous fluid.
In previous work, we sought to isolate the SGI by creating a set of
hydrostatic initial conditions such that the pressure gradient and the
acceleration due to self-gravity are both zero across the density interface.
In the present work, we relax this restriction and allow them
to be non-zero at the interface.  The self-gravitational acceleration can
effectively drive an RT-like instability provided the acceleration is 
in the same direction as the density gradient across the interface.
Although the self-gravitational acceleration varies with position, it may be
roughly constant near the interface where the growth is strongest. This RT-like
instability cannot be strictly isolated because the SGI always will 
arise in the presence of a density interface. Notwithstanding, the growth rate of 
the RT-like instability can be adjusted to grow faster than the SGI by increasing the
acceleration at the interface or decreasing the perturbation wavelength.

Section 2 describes the envisioned equilibrium configurations, while
\S 3 describes the two-dimensional computer simulations.  Section 4
discusses the results of our simulations, and \S 5 gives the conclusions
of this work.

\section{Equilibrium Configuration}
\begin{figure}
\includegraphics[scale=.6]{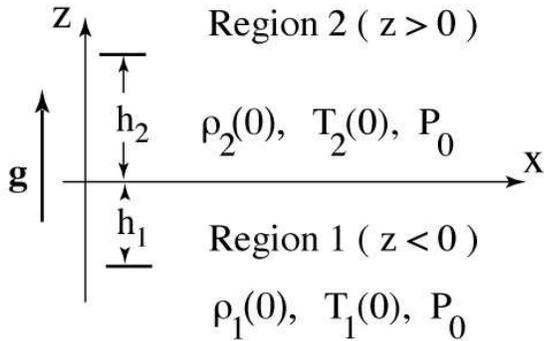}
\caption{Schematic of the equilibrium configuration.
\label{setup}}
\end{figure}

In order to isolate the gravitational instabilities, simulations are 
begun from a state of hydrostatic equilibrium. The geometry of the problem 
is shown in Figure \ref{setup}.
The interfacial values of the equilibrium densities and temperatures
in the two regions are $\rho_1(0) > \rho_2(0)$ and $T_1(0)<T_2(0)$, and the
interfacial pressure $P_0$ is the same for both media. The densities
and pressures lapse to zero at heights $h_1$ below and
$h_2$ above the interface. For RT simulations, a constant acceleration $g$ is
applied in the upward direction.

Adapting a polytropic equation of state ($P=\kappa \rho^\gamma$),
exact equilibrium solutions for a self-gravitating gas exist in the one
dimensional, unperturbed, planar problem when $\gamma$ = 2/3, 1, 2, and
$\infty$. A choice of $\gamma=2$ results in a stiffer equation of state
than is realistic for the ISM, but it admits simple, spatially bounded,
analytic expressions for the equilibrium distributions.
Previous studies revealed no differences in morphology or
growth rate between simulations with $\gamma=2$ and $\gamma=1.4$
\citep{h01}.
For $\gamma=2$, the hydrostatic and Poisson equations reduce to
\begin{equation}
\frac{d\rho_n(z)}{dz} = K_n f_n(z)~,
\end{equation}
and
\begin{equation}\label{poisson}
\frac{df_n(z)}{dz}= -4\pi G \rho_n(z)~.
\end{equation}
In the above, $n=1$ for region 1 and $n=2$ for region 2,
$K_n = [\rho_{n}(0)]^{2} / 2 P_0$, and $f_n(z)$ are the
self-gravitational accelerations. The solutions for the densities and
accelerations have the forms
\begin{equation}\label{rhoeqn}
\rho_n(z) = \rho_n(0) \cos\left(\frac{z}{l_n}\right) +
    \rho_n(0) \frac{f_0}{\beta} \sin\left(\frac{z}{l_n}\right)~,
\end{equation}
and
\begin{equation}\label{feqn}
f_n(z) = f_0\cos\left(\frac{z}{l_n}\right)
     -\beta \sin\left(\frac{z}{l_n}\right)~.
\end{equation}
The acceleration is continuous across the interface, $f_1(0) = f_2(0) = f_0$.
The constant $\beta$ is defined as
\begin{equation}
\beta = 2 \sqrt{2\pi G P_0}~.
\end{equation}
The gravitational scale heights obey the relation
\begin{equation}
l_n = \frac{1}{\rho_n(0)}\sqrt{\frac{P_0}{2\pi G}}~.
\end{equation}
Defining $h_1$ and $h_2$ as the absolute values of the heights at which
the density and pressure go to zero in the corresponding regions, the surface
densities are
\begin{equation}\label{sig2}
\sigma_2 \equiv \int^{h_2}_0 \rho_2(z)dz =
    \frac{\beta}{4\pi G}\sin\left(\frac{h_2}{l_2}\right) +
    \frac{f_0}{4\pi G} \left[1-\cos\left(\frac{h_2}{l_2}\right)\right]~,
\end{equation}
and
\begin{equation}\label{sig1}
\sigma_1 \equiv \int^{0}_{-h_1} \rho_1(z)dz =
    \frac{\beta}{4\pi G} \sin\left(\frac{h_1}{l_1}\right) -
    \frac{f_0}{4\pi G} \left[1-\cos\left(\frac{h_1}{l_1}\right)\right]~.
\end{equation}
Using equations (\ref{feqn}), (\ref{sig2}), and (\ref{sig1}),
it can be verified that Gauss' Theorem is satisfied,
\begin{equation}
f_2(h_2) - f_1(h_1) = -4\pi G(\sigma_2+\sigma_1)~.
\end{equation}
By symmetry, $f_2(h_2) = -f_1(h_1)$, which leads to the expression
\begin{equation}
f_2(h_2) = -2\pi G (\sigma_2 + \sigma_1)~.
\end{equation}
Upon integrating equation (\ref{poisson}) from $z=0$ to $z=h_2$,
we find
\begin{equation}
f_2(h_2) -f_0 = -4\pi G\sigma_2~,
\end{equation}
or
\begin{equation}
f_0 = 4\pi G\sigma_2 + f_2(h_2) =
4\pi G \sigma_2 - 2\pi G (\sigma_2 + \sigma_1)~,
\end{equation}
or
\begin{equation}
f_0 = 2\pi G (\sigma_2 - \sigma_1)~.
\end{equation}
Recalling that the densities in media 1 and 2 lapse to zero
at $z=-h_1$ and $z=h_2$, respectively, it follows from
equation (\ref{rhoeqn}) that
\begin{equation}
\tan \left( \frac{h_2}{l_2} \right) = -\frac{\beta}{f_0}~,
\end{equation}
and
\begin{equation}
\tan \left( \frac{h_1}{l_1} \right) = \frac{\beta}{f_0}~.
\end{equation}
Hereafter, we define $\theta_1 = h_1/l_1$ and
$\theta_2 = h_2/l_2$, both greater than zero.
The quadrants in which these angles are defined depends
upon the sign of $f_0$.  If $f_0$ is greater than zero,
$\theta_1$ is in the
first quadrant and $\theta_2$ is in the second quadrant.
Defining $\psi= \sqrt{\beta^2 + f_0^2}$, we have the relations
$\sin \theta_1 = \beta/\psi$, $\cos\theta_1 = f_0/\psi$,
$\theta_2 = \pi - \theta_1$, $\sin\theta_2= \sin\theta_1$,
and $\cos\theta_2= -\cos\theta_1$.
Therefore,
\begin{equation}
\sigma_2= \frac{\beta^2+f_0^2+f_0\psi}{4\pi G \psi}~,
\end{equation}
and
\begin{equation}
\sigma_1= \frac{\beta^2+f_0^2-f_0\psi}{4\pi G \psi}~.
\end{equation}
Consequently,
\begin{equation}
f_0=2\pi G (\sigma_2-\sigma_1) =
\frac{4\pi G f_0 \psi}{4\pi G \psi} = f_0~,
\end{equation}
an identity. The same process can be applied for the case 
$f_0<0$, with similar results. Therefore, the solutions form a consistent set.

In view of these results, we adopt the following strategy.
We set the molecular weight to $\mu=2$ for both media.
We specify the densities and temperatures at the interface
and calculate $P_0$, $\beta$, $l_1$, and $l_2$.
Then, we select $f_0$ and compute the density distributions
from equation (\ref{rhoeqn}).  The pressure and temperatures
distributions follow from the equation of state.  The final
step is to use equation (\ref{feqn}) to calculate the
boundary values of $f_n(z)$ for use by the gravity solver.

Due to discretization of the distributions across the grid, a truly
static state is not achieved.  We deem the setup to be sufficiently static if
the motions induced in the unperturbed case are negligible compared to any
imposed velocity perturbations. For example, a typical set of initial values 
at the interface is: $\rho_1(0)= 10 ^{-20}{\rm g ~cm}^{-3}$, $T_1(0) = 20 ~$K,
$\rho_2(0)= 0.2\times 10 ^{-20}{\rm g}\,{\rm cm}^{-3}$, and
$T_2(0) = 100 \, $K.
The adiabatic sound speed for a temperature of $20$ K
is $c = \sqrt{\gamma P_0 / \rho_1(0)} = 40,800\, {\rm cm}\,{\rm s}^{-1}$.
We typically use $5\%$ of the sound speed (about $2040~ {\rm cm}\,{\rm s}^{-1}$)
as the initial perturbation amplitude. If allowed to run to $5$ $e-$folding times
($1.06\times 10^{14}$s), the highest velocities observed
throughout the grid are less than $100\, {\rm cm}\,{\rm s}^{-1}$.
(This represents a conservative case; most static models show lower velocities.)
This is more than an order of magnitude lower than the
initial velocity perturbation amplitude. Deviations from the static solutions are
not large enough to affect the results of the perturbed simulations.

\section{Simulations}
As a test of the theory and for understanding the nonlinear behaviour 
of the instabilities we have carried out two-dimensional hydrodynamic 
simulations using CFDLib (Computational Fluid Dynamics Library), which 
was developed at the Los Alamos National Laboratory \citep{ka94}. 
The system of equations we solve is 
\begin{equation}\label{eq15}
\begin{array}{rcl}
\displaystyle{ \frac{\partial \rho}{\partial t} + {\bf \nabla}\cdot
\left( \rho {\bf v} \right)}& = &0~,\\[.5cm]
\displaystyle{\frac{\partial (\rho {\bf v})}{\partial t} + {\bf\nabla}\cdot
{\cal T} }& = &\rho ~{\bf g}~,\\[.5cm]
\displaystyle{   {\partial \over \partial t}
\left({1\over 2}\rho {\bf v}^2 +\rho \varepsilon \right)
+{\bf \nabla}\cdot \left[\left({1\over 2}\rho {\bf v}^2
+\rho w\right){\bf v}\right]} &=& 0 ~,\\[.5cm]
\displaystyle{\nabla \cdot {\bf g}} &= & 4\pi G \rho~,\
\end{array}
\end{equation}
where ${\cal T}_{jk}=p\delta_{jk}+\rho v_jv_k$ is the stress
tensor, ${\bf g}$ is the gravitational acceleration,
$\varepsilon$ is the specific energy of the fluid,
$w=\varepsilon+p/\rho$ is the enthalpy, and
the equation of state is $p=(\gamma-1)\rho \varepsilon$.
CFDLib is a finite-volume code well suited for problems of all flow speeds.
The self-gravitational potential is solved for
in two-dimensions using the MUDPACK multigrid code developed
at the National Center for Atmospheric Research (Adams 1989, 1991).
Models are run on a $2$D Cartesian mesh of size $257\times 257$.
Simulations repeated on a $513\times 513$ grid show a
difference in fine scale structure but not in growth rate. The normal velocity 
components of the gas are confined by reflective boundary conditions 
on all sides; whereas, the gravitational potential solver uses
periodic boundary conditions along the side boundaries and specified
gradient conditions along the top and bottom.

We induce perturbations along the interface through a velocity function of the form
\begin{equation}
v(x,z) = v_0 \cos(kx)\exp(-k |{z}|)~,
\end{equation}
with $v_0$ set to $ 5\%$ of the sound speed in the denser fluid. The perturbation 
is localized to the interface due to the exponential factor in the vertical, 
$z$-direction.  We use a velocity perturbation instead of a spatial perturbation
for two reasons. While our grid resolution is sufficient to determine growth
rates and obtain a sensible picture of nonlinear structure, it is too coarse to
impose a spatial perturbation without giving rise to spurious instabilities due to
the square cell structure. Also, without careful consideration,
imposing a perturbation across an interface gives rise to a decaying as well
as a growing mode. The effect of the decaying mode upon the velocity is easily
seen and considered in determining growth rates.

For ease of notation, we quote all times in units of the $e$-folding  time for the
incompressible, linear SGI ($g=0$) as determined from equation (\ref{rates}),
$t_e=\omega^{-1} =  2.11\times{10}^{13}$s.
We normalize all growth rates by dividing by the corresponding SGI growth rate,
$\omega_{SGI}= t_e^{-1} = 4.73\times 10^{-14}{\rm s}^{-1}$.
We define two ratios for each model:
$\Omega_S$ for the growth of dense spikes, and $\Omega_B$ for the growth of 
tenuous bubbles. We can define an average normalized
growth rate $\Omega_T=0.5(\Omega_B+\Omega_S$).
This language is consistent with typical RT descriptions.

For this study, we define two different classes of models.
For the first suite of models, we selected a perturbation
wavelength, $\lambda = 3.21\times 10^{17}$cm, a value nearly a factor of ten
smaller than the Jeans length in the denser fluid.
We estimate the acceleration that would give rise to an RT instability 
with the same growth rate as the SGI,  $g= \sqrt{2} G (\rho_1(0) - \rho_2(0)) 
\lambda$ = $2.43\times 10^{-10}{\rm cm}\,{\rm s}^{-2}$.
(The factor $\sqrt{2}$ is a geometric correction
to account for the use of a one-dimensional wave vector instead of a
two-dimensional wave vector. The value of $k$ is reduced by $\sqrt{2}$ in going
from a $2$D to a $1$D wavevector, so the value of $g$ must increase by 
$\sqrt{2}$ to arrive at the correct growth rate.
By the same logic, we also adjust our definition of the Jeans wavelength in
\S 4.3 to preserve the relationship $k\lambda=2\pi$.)
We compare the rate of perturbation growth for accelerated ($f_0 \ne 0$) interfaces
against both the pure SGI ($f_0=0$) and RT ($g>0$, no self-gravity) cases.
Our strategy is to select $f_0$ values that should lead to growth rates 
equal to $q\omega_{SGI}$, where $q=1$, $2$, and $3$.
We compare the theoretical normalized growth rates $q$ to the
calculated values $\Omega_T$. In addition, we look at models having
$f_0 = -g \pm 1\times 10^{-10} {\rm cm}\,{\rm s}^{-2}$, in order to ascertain the critical
$f_0$ value defining the boundary between stable and unstable behaviour.

We follow with a second set of models designed to investigate the SGI as
perturbation wavelengths approach the Jeans length. We remove the RT
component ($f_0=g=0$) and use a value $\gamma=1.1$ to allow more compression.
When $f_0=0$, the initial distributions in density, temperature, and acceleration
can be determined numerically for any value of $\gamma$ (Hueckstaedt 2001).
At question is whether the SGI can drive a system which is otherwise marginally
Jeans stable toward global collapse.

\section{Results}
\begin{figure}
\includegraphics[scale=.5]{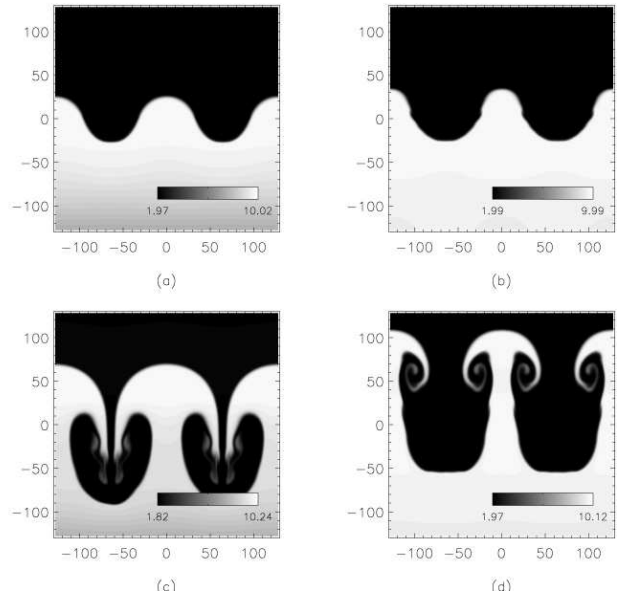}
\caption{Density contours for the SGI [(a) and (c)] and the
RT instability [(b) and (d)] for $t=2t_e$ and $t=4t_e$.
The grey-scales show the density values $\times{10}^{-21}$.
The $e-$folding time $t_e\!=\!2.11\times{10}^{13}\,$s is determined
from equation \ref{rates} and is used for all figures.
\label{sgirt} }
\end{figure}

\begin{figure}
\includegraphics[scale=.6]{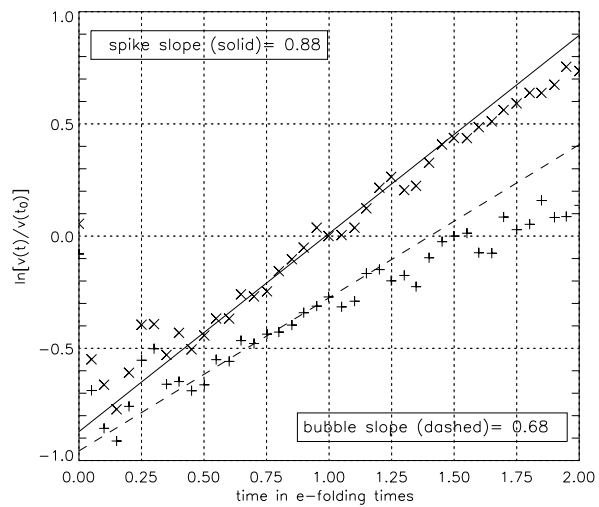}
\caption{Velocity growth for RT instability.
The maximum spike ($\times$) and
bubble ($+$) velocities are plotted versus time using different
values for the normalization velocity.  Lines are drawn to represent the
linear phase growth, with the slopes determining the growth rates.
\label{rtrate} }
\end{figure}

In Papers 3 and 4, we compared and contrasted the growth of pure SGI and
RT instabilities. An example of previous results is shown in Figure
\ref{sgirt}.  Density contours for SGI and RT models with identical
growth rates as determined by equation \ref{rates} are plotted for times of $2t_e$ and
$4t_e$. As is characteristic of the SGI, the tenuous fingers grow more rapidly
than the dense spikes. All of the models presented in this communication
share the same theoretical $e-$folding time for a pure SGI instability,
$t_e=\omega^{-1} =  2.11\times{10}^{13}$s. All times are quote in units of $t_e$.
Figure \ref{rtrate} shows a typical velocity plot used to determine numerical
growth rates. The logarithmic values of the maximum velocity in the spikes 
and bubbles are plotted for the RT instability.
Lines are fit to the velocity points and slopes determined to represent
the linear growth rates of both spikes and bubbles. Numerical growth rates for
both the RT instability and SGI do not exactly match theoretical values.
Rather, the ratio of computational to theoretical growth rate varies with
velocity perturbation amplitude and density ratio across the interface (Paper
4).

\subsection{Modifying growth rates with non-zero $f_0$}
In order to generalize our study, we now allow non-zero values for $f_0$.
However, we do so without an imposed constant acceleration (i.e. no $g$ term).
So the RT-like contribution to the perturbation growth comes solely  from the
self-gravitational term at the interface. For $f_0>0$, an upward acceleration 
drives the interface in a RT-unstable manner.  If $f_0<0$, the RT-like 
component is stabilizing, but since the SGI is always destabilizing, 
perturbations may still grow if the RT-like term is relatively small.  
We investigate the two  issues with the set of models summarized in Table 1.
\begin{table}
\caption{Model Parameters}
\begin{tabular}{lcccccc}
\hline
model & $f_0$ & $q$ & $\Omega_S$ & $\Omega_B$ & $\Omega_T$ &Figure \#\\
\hline
f0  & 0.0      & 1 & 0.773 & 0.883 & 0.828 & 2\\
f3  & 7.277  & 2 & 1.800 & 1.529 & 1.665 & 6\\
f8  & 19.41 & 3 & 3.090 & 2.646 & 2.868 & 9\\
\hline
deq & -2.426 & 0  &  ---  & ---   & ---   & 11 \\
dpl & -1.426 & -- &  ---  & ---   & ---   & -- \\
dmi & -3.428 & -- &  ---  & ---   & ---   & 12 \\
\hline
\end{tabular}

Units for Table 1:
$f_0 \rightarrow 10^{-10}{\rm cm}\,{\rm s}^{-2}$
\end{table}
First, we examine the overall growth rates when an unstable RT-like
term is added to the SGI at the interface. Next we examine the marginal case
where $f_0=-g$ along with a small deviation in each direction.

\begin{figure}
\includegraphics[scale=.45]{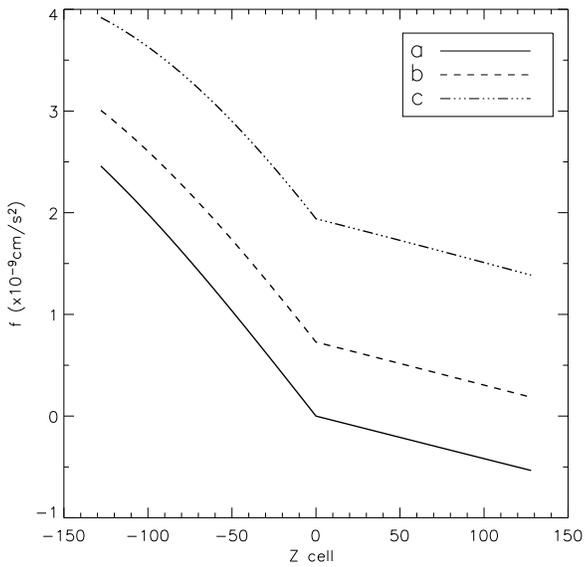}
\caption{Acceleration due to self-gravity ($f$) for
(a) model f0: $f_0=0$,(b) model f3:
$f_0=7.28\times10^{-10}{\rm cm}\,{\rm s}^{-2}$,
and (c) model f8: $f_0=1.94\times10^{-9}{\rm cm}\,{\rm s}^{-2}$.
\label{plotf} }
\end{figure}

\begin{figure}
\includegraphics[scale=.45]{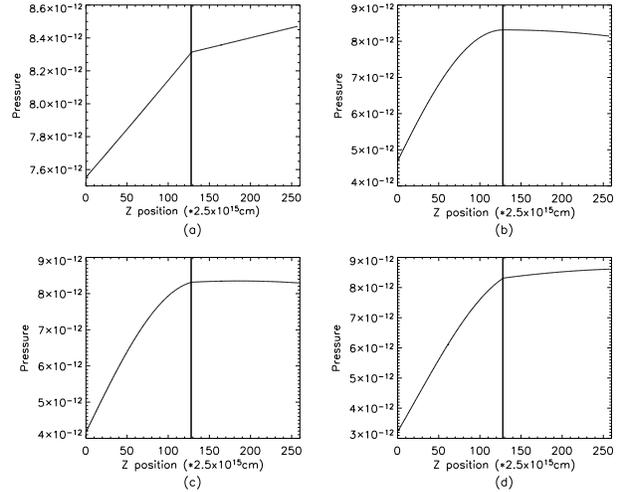}
\caption{Initial pressure distributions for (a) RT instability with
$g= 2.43\times10^{-10}{\rm cm}\,{\rm s}^{-2}$, (b) model f0,
(c) model f3, and (d) model f8.
The vertical lines indicate the interface locations.
\label{x4pr} }
\end{figure}

The first study consists of three models: model f0 for $f_0=0$, the pure SGI; model f3
for $f_0=7.28\times 10^{-10}{\rm cm}\,{\rm s}^{-2}$, which is three times the 
value of $g$ that would result in a pure RT instability with the same theoretical
linear growth rate as model f0; and model f8 for $f_0=19.4\times 10^{-10}
{\rm cm}\,{\rm s}^{-2}$, with eight times the canonical value of $g$.
The accelerations due to self-gravity at time zero
for these three models are shown in Figure \ref{plotf}.
A higher value of $f_0$ results in higher accelerations throughout the grid.
The resultant equilibrium pressure distributions for these models (plus the pure
RT instability) are shown in Figure \ref{x4pr}.
For the pure SGI, the pressure falls in both directions away from the interface.
But as $f_0$ is increased, the acceleration in the less dense medium is
positive instead of negative. As a result, a monotonically increasing pressure is
required to form an equilibrium.

\begin{figure}
\includegraphics[scale=.5]{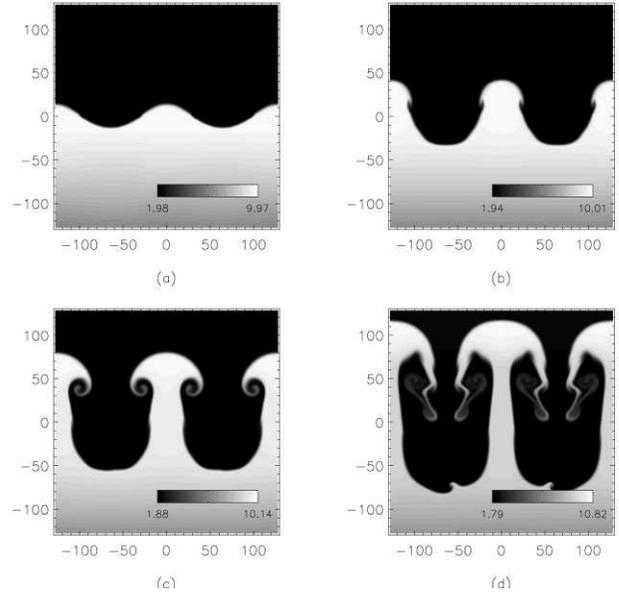}
\caption{Density contours for the SGI model f3 at
(a) $0.9t_e$, (b) $1.5t_e$, (c) $2.1t_e$, and (d) $2.7t_e$.
\label{x3p2} }
\end{figure}

\begin{figure}
\includegraphics[scale=.45]{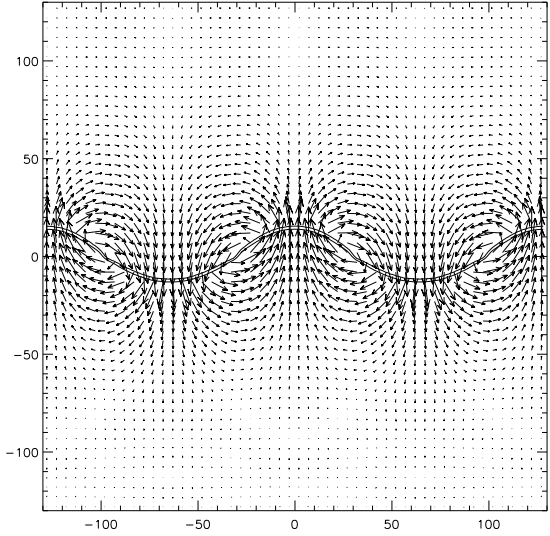}
\caption{Velocity map for model f3 at $t=0.9t_e$. Arbitrary 
density contours are overplotted  to show the interface position.
\label{x3p2vel1} }
\end{figure}

\begin{figure}
\includegraphics[scale=.45]{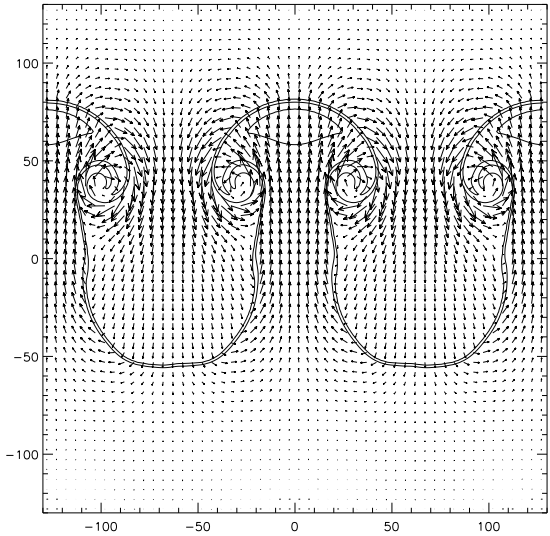}
\caption{Velocity map for model f3 at $t=2.1t_e$.
\label{x3p2vel2} }
\end{figure}

The values of $f_0$ for models f3 and f8 are chosen such that by equation
(\ref{rates}) we expect twice and three times the growth rate, respectively,
of model f0.  In Figure \ref{x3p2} we show the evolution of model f3.
Recall that the times are quoted in units of the $e-$folding time for model f0.
A comparison of figures \ref{x3p2} and \ref{sgirt} shows that a non-zero
value of $f_0$ does indeed lead to a hybrid sort of structure.
The mushroom caps seen at later times are broader for model f3 than those in the pure
RT case but slimmer than those in model f0. Also, the Kelvin-Helmholtz
roll-ups in model f3 have assumed a more relaxed shape than the tight
rolls seen in the pure RT. Velocity vectors are plotted for
model f3 in Figures \ref{x3p2vel1} and \ref{x3p2vel2} at different times to
illustrate the circulation patterns that develop.

\begin{figure}
\includegraphics[scale=.5]{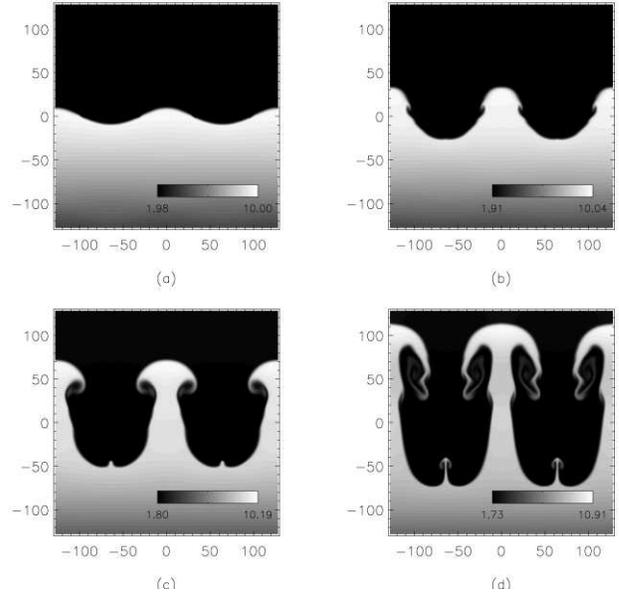}
\caption{Density contours for the SGI model f8 at
(a) $0.6t_e$, (b) $1.0t_e$, (c) $1.4t_e$, and (d) $1.8t_e$.
\label{x8p2} }
\end{figure}

\begin{figure}
\includegraphics[scale=.6]{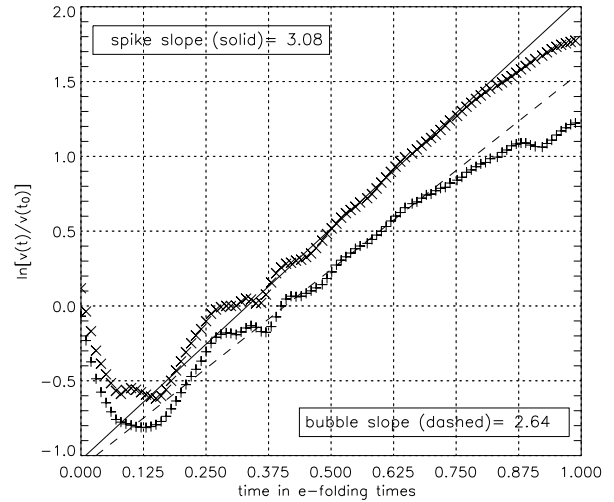}
\caption{Velocity growth for model f8.  The spike and bubble velocities are
plotted with different normalization values.
\label{x8rate} }
\end{figure}

The hybrid structure persists in model f8 (Figure \ref{x8p2}).
The mushroom caps are nearer in size than those in
model f3 to the pure RT case, but the Kelvin-Helmholtz roll-ups 
retain their SGI-like relaxed shape.
As listed in Table 1, both f3 and f8 have larger linear growth
rates for the dense spikes than the bubbles ($\Omega_S>\Omega_B$).
This is shown graphically for model f8 in Figure \ref{x8rate}.
In this respect, the RT-like growth induced by the relatively large values for
$f_0$ dominate over the SGI component during the early phases of growth.
For all three models, the calculated average growth rate $\Omega_T$ is a bit lower
than the predicted $q$ from simplified, incompressible theory.
The calculated growth rate for model f3 is twice that for model f0, as expected.
However, the calculated growth rate for model f8 is relatively high at nearly
$3.5$ times that of model f0. This larger than expected growth rate may be an
indication that the amplified RT component grows too fast to be affected by the SGI
component of the growth. We note that the calculated growth rate for spikes in
model f8 ($\Omega_S=3.09$) is three times the expected spike growth rate for a
pure RT instability with $g=f_0$. Thus neither the SGI, nor deviations from
incompressibility, appear to alter the rapid RT-like growth when $f_0$ is very
large.

\begin{figure}
\includegraphics[scale=.5]{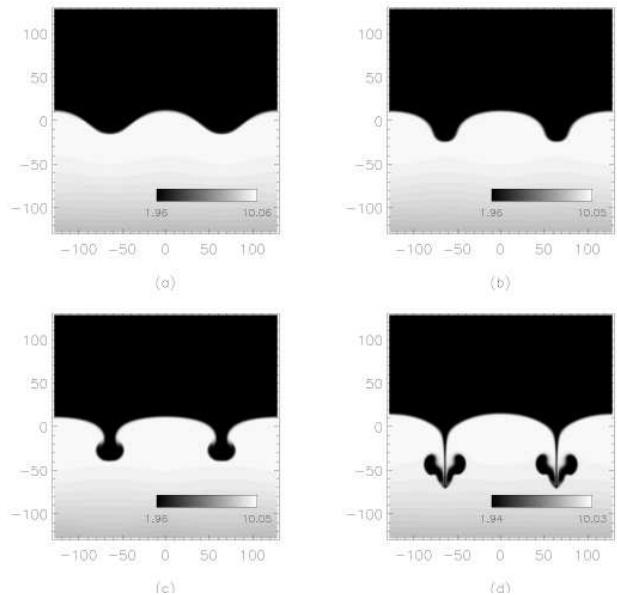}
\caption{Density contours for the SGI with
$f_0 = -g = -2.43\times10^{-10}{\rm cm}\,{\rm s}^{-2}$ at
(a) $2.0t_e$, (b) $3.0t_e$, (c) $4.0t_e$, and (d) $5.0t_e$.
\label{x1p2} }
\end{figure}

\begin{figure}
\includegraphics[scale=.5]{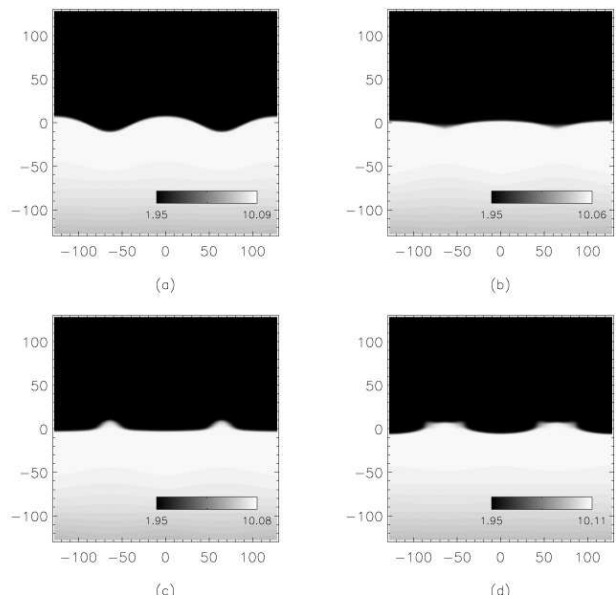}
\caption{Density contours for the SGI with
$f_0 = -3.43\times10^{-10}{\rm cm}\,{\rm s}^{-2}$ at
(a) $2.0t_e$, (b) $3.0t_e$, (c) $4.0t_e$, and (d) $5.0t_e$.
\label{x1mp2} }
\end{figure}

We now ask what happens if we apply a negative value of $f_0$. In this case,
$f_0(\rho_1-\rho_2)<0$ so the RT-like component is stabilizing.
The evolution for model deq (Figure \ref{x1p2}),
for which $f_0=-g$, is not quite stable even though incompressible
theory predicts $q=0$. The structure resembles the growth of the SGI more
than that of the RT instability.
The inherently unstable SGI seems more robust than the stabilizing RT component.
If the RT component is weakened (decreased in absolute value), as in model dpl, the
result is unchanged except for a slight increase in the rate of growth.
If the RT component is increased in absolute value
(model dmi), the SGI growth is quenched (Figure \ref{x1mp2}).

\subsection{Long wavelength perturbations with $\gamma=1.1$}
For another set of models, we decreased the ratio of specific heats
and set $f_0=0$. We seek to examine the behaviour of the SGI as
perturbation wavelengths approach the Jeans wavelength for
gravitational collapse in the relatively dense, cool gas of medium 1 at the interface.
We recall that the Jean wavelength ($\lambda_J$) is the wavelength at which 
the force of self-gravity exactly balances the stabilizing pressure force 
in an infinite, uniform, isothermal medium. Adjusting for non-unity 
$\gamma$, the three-dimensional Jeans length is given by
$\lambda_{J3d} = \rho_{1}^{-1}(0)\sqrt{(\pi\gamma P_0)/G} =
2.07\times10^{18}$cm. We multiply this result by $\sqrt{2}$ to
account for the two-dimensional nature of our simulations and arrive at the value
$\lambda_J = 2.92\times10^{18}$cm. In an idealized model, no 
gravitational collapse is expected when $\lambda=\lambda_J$; 
a significantly longer wavelength is required for collapse to occur.
In contrast, the SGI grows for all wavelengths, generating
velocities which can lead to large compression if given an additional
gravitational boost.

\begin{figure} %%%%%%%%%
\includegraphics[scale=.5]{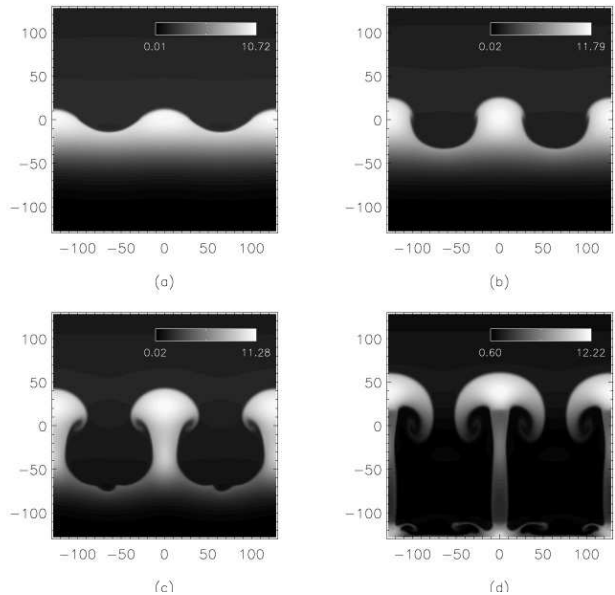}
\caption{Density contours for the SGI model LJ2 with
$\lambda = \lambda_J/2$ at
(a) $3.0t_e$, (b) $4.0t_e$, (c) $5.0t_e$, and (d) $6.0t_e$.
\label{jeans2p2} }
\end{figure}

\begin{figure} %%%%%%%%%%
\includegraphics[scale=.45]{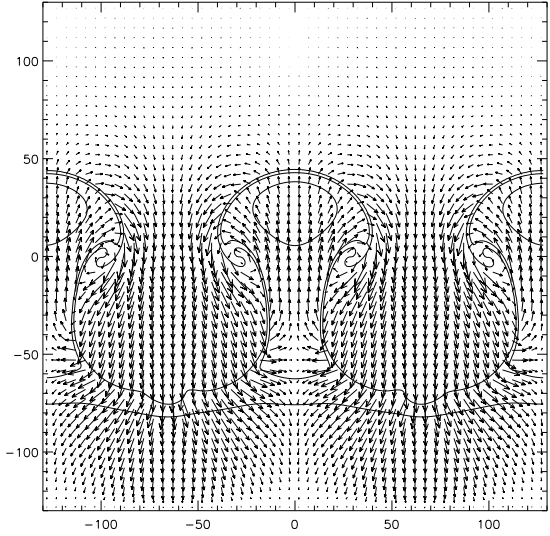}
\caption{Velocity map for model LJ2 at $t=5.0t_e$.
\label{veljeans} }
\end{figure}

\begin{figure} %%%%%%%%%%%%
\includegraphics[scale=.5]{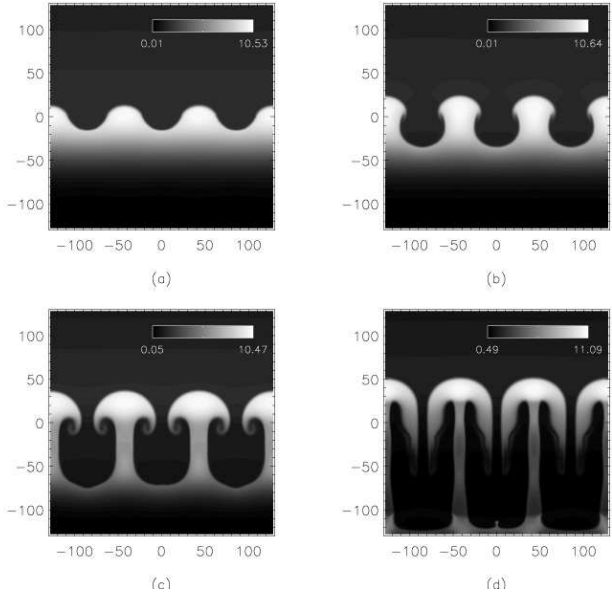}
\caption{Density contours for the SGI model LJ3 with
$\lambda =  \lambda_J/3$ at (a) $3.0t_e$, (b) $4.0t_e$, (c) $5.0t_e$, 
and (d) $6.0t_e$.
\label{jeans2p3} }
\end{figure}

Other than the change to $\gamma=1.1$, the physical parameters for the three
models listed in Table 2 are identical to model f0. We increase the cell size in
both directions to $\Delta{x}=\Delta{y}= 1.1373\times10^{16}$cm, so that the
placement of two waves across $257$ cells results in a perturbation wavelength
$\lambda=\lambda_J/2$ (model LJ2). The evolution of this model is shown
in Figure \ref{jeans2p2}.
\begin{table}
\caption{Model Parameters}
\begin{tabular}{lcccc}
\hline
model & ${\lambda}/{\lambda_J}$ & $\rho_{\max}(3t_e)$ &
                                  $\rho_{\max}(6t_e)$ & Figure \#\\
\hline
LJ1  & 1  &  10.8 & 20.4 & 16\\
LJ2  & 1/2 & 10.7 & 12.2 & 13\\
LJ3  & 1/3 & 10.5 & 11.1 & 15\\
\hline
\end{tabular}

$\lambda_J=2.92\times10^{18}$cm; densities given in units of
$10^{-21}{\rm g}\,{\rm cm}^{-3}$
\end{table}
The larger cell size results in poorer resolution of the roll-up features. This
is the price paid to be able to examine large scale collapse behaviour without
greatly increasing the computational cost. Velocity vectors for model LJ2 are
shown in Figure \ref{veljeans} for $t=5t_e$.
By placing three waves across the $257$ cell extent of the grid, we arrive
at model LJ3 with $\lambda=\lambda_J/3$ (Figure \ref{jeans2p3}).
As expected, the shorter wavelength of model LJ3 results in a growth rate and
morphology similar to those for model LJ2.

\begin{figure}  %%%%%%%%%%%
\includegraphics[scale=.5]{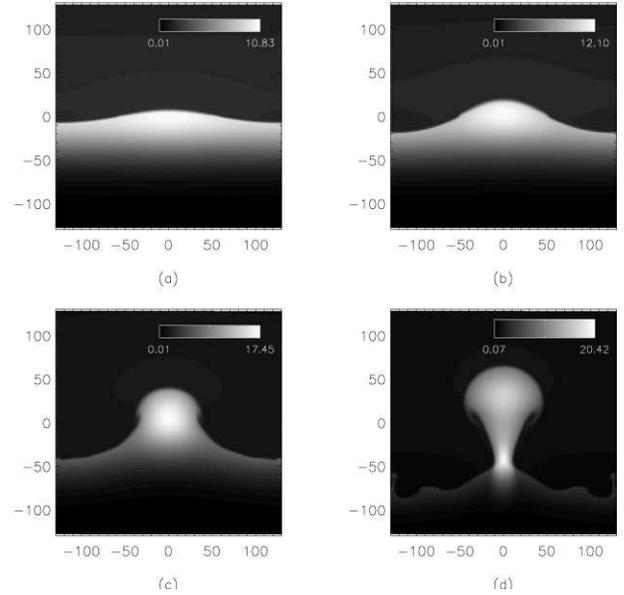}
\caption{Density contours for the SGI model LJ1 with
$\lambda = \lambda_J$ at
(a) $3.0t_e$, (b) $4.0t_e$, (c) $5.0t_e$, and (d) $6.0t_e$.
\label{jlam} }
\end{figure}

\begin{figure}
\includegraphics[scale=.5]{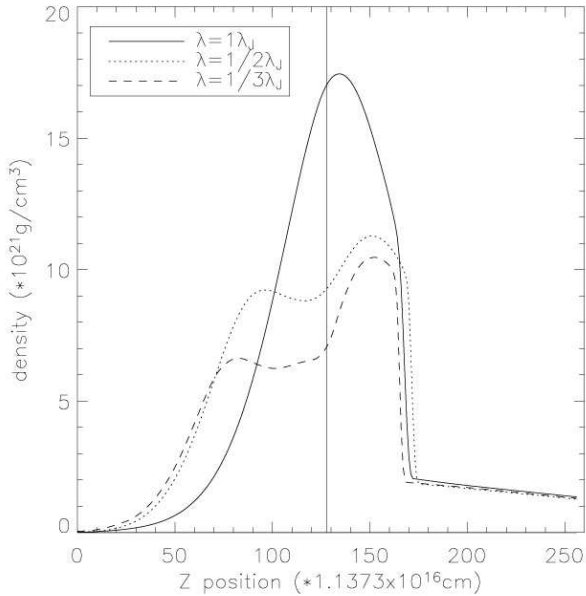} %%%%%%%%%%%%
\caption{Density cuts through columns at $t=5.0t_e$ for three different
wavelengths. The vertical line indicates the initial location of the interface.
\label{rhocut5} }
\end{figure}

\begin{figure}  %%%%%%%%%%%%%
\includegraphics[scale=.5]{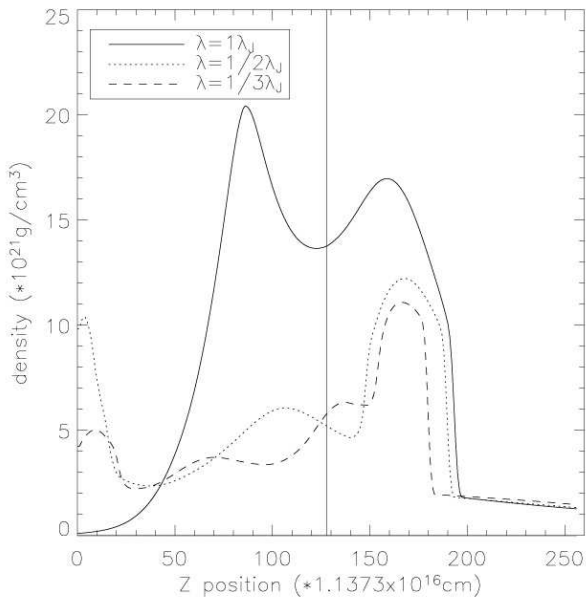}
\caption{Density cuts through columns at $t=6t_e$ for three different
wavelengths.
\label{rhocut6} }
\end{figure}

We observe different behaviour when $\lambda=\lambda_J$ (model LJ1).
For this model, we extended the computational grid in the  $x$ 
direction to $513$ cells and maintained the same cell size.
As shown in Figure \ref{jlam}, the downward motion of the tenuous 
bubbles (which outpaces the upward moving spikes in models LJ2 and LJ3) 
is slower  than observed for the other models. Only the central part of the 
grid is shown in order to maintain a visual scale consistent with the 
previous figures.
\begin{figure} %%%%%%%%%%%%%
\includegraphics[scale=.45]{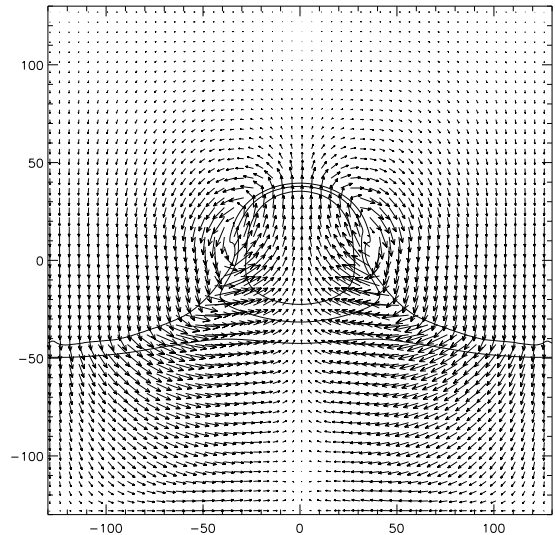}
\caption{Velocity map for model LJ1 at $t=5.0t_e$.
\label{veljlam} }
\end{figure}
Velocity vectors for model LJ1 are shown in Figure \ref{veljlam} for $t=5t_e$.
Although the maximum velocities appearing in Figures \ref{veljeans} and
\ref{veljlam} are comparable (about 0.4 km/s), model LJ1 exhibits a greater
degree of collapse as indicated by the converging velocity vectors.
The collapsing nature of model LJ1 is highlighted by taking density line-outs
through dense columns for all three models at $t=5t_e$ (Figure \ref{rhocut5})
and $t=6t_e$ (Figure \ref{rhocut6}).
The densest regions in models LJ2 and LJ3 are found in the dense mushroom caps
which move upward over time. In model LJ1, the densest features form near the
interface.  At $t=6t_e$, the highest compression is seen to have  occurred at the 
stem of the spike rather than in the cap. We truncate our analysis at $t=6t_e$
due to boundary effects from the top and bottom of the computational grid.
Also, we consider the cylindrical symmetry of our two-dimensional calculations
unrealistic after the onset of (three-dimensional) collapse.
As an estimate of the minimum collapsing mass, we note that the mass 
of a sphere of diameter $\lambda_J/2$ (the positive phase of the perturbation for
model LJ1) and uniform density $\rho_{1}(0)$ is about $8.2M_\odot$.

\section {Conclusions}

Based upon the work summarized here, we conclude that
large scale, planar density discontinuities (or near discontinuities) are
inherently unstable on a gravitational collapse timescale in two respects:
1. Any density interface is inherently unstable to the SGI with a
growth rate that is scale invariant for incompressible media and nearly scale invariant
compressible media. (In compressible media, some of the energy
goes into compression at the expense of driving fluid displacement.)
2. In general. the acceleration due to self-gravity will be non-zero at a
density interface. which will add a Rayleigh-Taylor component to the
instability if $f_0(\rho_1-\rho_2)>0$. These instabilities cause crenulations
along an interface to grow into spikes and bubbles with a growth rate
and nonlinear structure that depend upon the relative strength of the SGI 
and RT-like components. For wavelengths that approach and exceed the 
Jeans length in the denser medium, structures initiated by interfacial 
instabilities are likely to undergo continued gravitational collapse.

An inspection of Hubble Telescope images of interstellar clouds 
reveals  crenulated interfaces such as the ``elephant trunks'' and 
``pillars'' in the Eagle Nebula  (Hester et al. 1996; Pound 1998). 
These structures have commonly been interpreted in terms of the 
Rayleigh-Taylor instability  of an ionization front created by the UV radiation
from nearby OB stars, but without any consideration of the self-gravity of the cloud.
However, recent simulations of the ionization front dynamics indicate that the
Rayleigh-Taylor instability is quenched when hydrogen recombination is included 
(Mizuta et al. 2005) as predicted by Kahn (1958).
It is clearly of interest to investigate the ionization front stability including self-gravity
to see if the SGI instabiliy overcomes the stabilizing effect of recombination. 
For this we would need to modify the third line of equation (20) to 
include the different energy sources and losses due to the UV absorption,
recomination, and radiative cooling. It is also of interest to extend our work to
larger computational grids which will allow us to follow perturbation
growth to longer times. A more ambitious objective is to simulate
the SGI in three-dimensions at both short and long wavelengths.

Gravitational processes (both global and interfacial) clearly have an important role in the evolving ISM.
For example, Burkert and Hartmann (2004) have shown that gravitational
forces give rise to a variety of structures along the edges of finite, self-gravitating
sheets in a manner consistent with observations of local molecular clouds.
In the final stages of star formation, gravitational forces dominate.
But  long before the final  stages of collapse, gravitational forces are 
important for driving instabilities which convert gravitational 
energy into flow kinetic energy, enhance density inhomogeneities, and determine
the partitioning of energy between different length scales.

\section* {acknowledgments}
The authors thank B.A.\ Kashiwa and N.T.\ Padial for help in
modifying CFDLib for astrophysical use.
The work of R.L.  was partially supported by CRDF
grant KP2-2555-AL-03.
Los Alamos National Laboratory is operated by the University of California
for the U.S.\ Department of Energy under contract No.\ W-7405-Eng-36.

%%%%%%%%%%%%%%%%%%%%%%%%%%%%%%%%%%%%

% \bsp % ``This paper has been produced using the ...''

\label{lastpage}

\end{document}